\documentclass[twocolumn,english,prb,aps,final,superscriptaddress]{revtex4}
\usepackage[latin9]{inputenc}
\usepackage{xcolor}
\usepackage{babel}
\usepackage{textcomp}
\usepackage{amsmath}
\usepackage{amssymb}
\usepackage{graphicx}
\usepackage[unicode=true,
 bookmarks=true,bookmarksnumbered=true,bookmarksopen=true,bookmarksopenlevel=1,
 breaklinks=true,pdfborder={0 0 1},backref=false,colorlinks=true]
 {hyperref}
\hypersetup{pdftitle={Dielectric response of ferroelectric relaxors by statistical modeling},
 pdfauthor={D. Wang},
 pdfsubject={physics},
 pdfkeywords={relaxor, ferroelectircs, statistics, Maxwell-Boltzmann distribution}}

\makeatletter

\providecommand{\tabularnewline}{\\}

\@ifundefined{textcolor}{}
{%
 \definecolor{BLACK}{gray}{0}
 \definecolor{WHITE}{gray}{1}
 \definecolor{RED}{rgb}{1,0,0}
 \definecolor{GREEN}{rgb}{0,1,0}
 \definecolor{BLUE}{rgb}{0,0,1}
 \definecolor{CYAN}{cmyk}{1,0,0,0}
 \definecolor{MAGENTA}{cmyk}{0,1,0,0}
 \definecolor{YELLOW}{cmyk}{0,0,1,0}
}

\@ifundefined{date}{}{\date{}}

\usepackage{graphics}

\newcommand{\mathsym}[1]{{}}
\newcommand{\unicode}[1]{{}}

\usepackage{times}
\usepackage{amsmath}

\usepackage{mathptmx}

\makeatother

\begin{document}

\title{Insight into dielectric response of ferroelectric relaxors by statistical
modeling}

\author{J. Liu}

\affiliation{State Key Laboratory for Mechanical Behavior of Materials, School
of Materials Science and Engineering, Xi\textquoteright an Jiaotong
University, Xi'an, China, 710049}

\author{F. Li }

\affiliation{Electronic Materials Research Laboratory-Key Laboratory of the Ministry
of Education and International Center for Dielectric Research, Xi'an
Jiaotong University, Xi'an 710049, China}

\author{Y. Zeng}

\affiliation{School of Materials Science and Engineering, State Key Laboratory
of New Ceramics and Fine Processing, Tsinghua University, No.1, Qinghua
Yuan, Beijing, 100084, China}

\author{Z. Jiang }

\affiliation{School of Electronic and Information Engineering \& State Key Laboratory
for Mechanical Behavior of Materials, Xi\textquoteright an Jiaotong
University, Xi\textquoteright an 710049, China}

\affiliation{Physics Department and Institute for Nanoscience and Engineering,
University of Arkansas, Fayetteville, Arkansas 72701, USA }

\author{D. Wang}
\email{dawei.wang@mail.xjtu.edu.cn}

\selectlanguage{english}%

\affiliation{School of Electronic and Information Engineering \& State Key Laboratory
for Mechanical Behavior of Materials, Xi\textquoteright an Jiaotong
University, Xi\textquoteright an 710049, China}

\author{Z.-G. Ye}

\affiliation{Department of Chemistry and 4D LABS, Simon Fraser University, Burnaby,
British Columbia, Canada V5A 1A6}

\affiliation{Electronic Materials Research Laboratory-Key Laboratory of the Ministry
of Education and International Center for Dielectric Research, Xi'an
Jiaotong University, Xi'an 710049, China}

\author{C.-L. Jia}

\affiliation{School of Electronic and Information Engineering \& State Key Laboratory
for Mechanical Behavior of Materials, Xi\textquoteright an Jiaotong
University, Xi\textquoteright an 710049, China}

\affiliation{\textsuperscript{}Peter Grünberg Institute and Ernst Ruska Center
for Microscopy and Spectroscopy with Electrons, Research Center Jülich,
D-52425 Jülich, Germany}

\date{\today}
\begin{abstract}
Ferroelectric relaxors are complex materials with distinct properties.
The understanding of their dielectric susceptibility, which strongly
depends on both temperature and probing frequency, have interested
researchers for many years. Here we report a macroscopic and phenomenological
approach based on statistical modeling to investigate and better understand
how the dielectric response of a relaxor depends on temperature. Employing
the Maxwell-Boltzmann distribution and considering temperature dependent
dipolar orientational polarizability, we propose a minimum statistical
model and specific equations to understand and fit numerical and experimental
dielectric responses versus temperature. We show that the proposed
formula can successfully fit the dielectric response of typical relaxors,
including Ba(Zr,Ti)O$_{3}$, Pb(Zn$_{1/3}$Nb$_{2/3}$)$_{0.87}$Ti$_{0.13}$O$_{3}$,
and Pb(Mg$_{1/3}$Nb$_{2/3}$)O$_{3}$-0.05Pb(Zr$_{0.53}$Ti$_{0.47}$)O$_{3}$,
which demonstrates the general applicability of this approach. 
\end{abstract}
\maketitle

\section{Introduction}

Relaxor ferroelectrics are materials that exhibit interesting dielectric
responses different from normal ferroelectrics. For instance, they
often possess relaxation modes at low frequency ($<1$\,GHz). Relaxors
ferroelectrics have been exploited in many applications such as actuators
due to their giant electromechanical couplings \cite{Uchino1996},
and their properties extensively investigated, including structural
properties (e.g., polar nanoregions or PNRs) using neutron scattering
\cite{PhononLocalization}, dielectric responses \cite{Bokov2006a,Nuzhnyy2012,Petzelt2014,Kleeman2014,Wang2014,Wang2016},
the crossover from ferroelectrics to relaxor \cite{Kleeman2014}.
To understand such systems, many theories have been proposed \cite{Pirc1999,Bokov2012,Akbarzadeh2012,Sherrington2013a,Uchino2014,Sherrington2014,PhononLocalization,Kleemann2015}.
Relaxors are complex systems, to some extent similar to spin glasses
\cite{Sherrington2013a,Sherrington2014a,Sherrington2014}, in that
their compositions are, without exception, made of complex oxides
containing different ions and inevitably inhomogeneous. For instance,
the B-site ions of typical relaxors Ba(Zr,Ti)O$_{3}$ (BZT) and Pb(Mg$_{1/3}$Nb$_{2/3}$)O$_{3}$(PMN)
are randomly distributed.

The dielectric response of ferroelectric relaxors is the defining
feature that differentiates them from normal ferroelectrics: (i) large
susceptibilities at low frequency (GHz or lower) ; (ii) even more
unusual, the characteristic temperature $T_{m}$, at which the susceptibility
peaks strongly depends on the frequency of the probing \emph{ac} electric
field. In other words, the susceptibility, $\chi$, depends on both
temperature, $T$, and the probing frequency, $\nu$. While such phenomena
are well known experimentally \cite{Bokov2007BZT,Bokov2012,Gridnev2004,Bokovo2006,BZT-susceptibility,Tagantsev1999,PMN-BZT},
numerical generation of relaxor's dielectric response from model-based
simulations has been a challenging work. For instance, the shift of
$T_{m}$ of the lead free relaxor BZT was only numerically achieved
recently\cite{Wang2016}. Since numerous ferroelectric relaxors exist,
numerically treating each of them remains a daunting task. \,\,\,One
way to mitigate this difficulty is to resort to statistical modeling
\cite{Sethna2010}. For a complex system, a statistical approach can
provide intuitive understanding by capturing dominant factors, provide
equations to understand experimental results. and help extracting
useful information. In the present work, we adopt this approach to
treat the dielectric response of relaxors and show that such a statistical
model can indeed be applied to understanding how the dielectric constants
change with temperatures and probing frequency.

While the susceptibility of relaxors, $\chi\left(T,\nu\right)$, depends
on both temperature and frequency, theoretical models are often proposed
to treat $\nu$ and $T$ separately \cite{Uchino2014,Bokov2012,Cross1987a,UniversalRelaxation,DielRelaxSolids,Nuzhnyy2012}.
For instance, at a given temperature, two processes are employed in
the fitting of $\chi\left(\nu\right)$ of Ba (Ti$_{0.675}$Zr$_{0.325}$
)O$_{3}$\textcolor{black}{{} \cite{Bokov2007BZT,BZT-susceptibility}}:
the universal relaxor process and the conventional relaxor process,
which have different relaxation characteristics employing the Curie-Von
Schweidler law at low frequency and the Kohlrausch-Williams-Watts
law at higher frequency \cite{Bokov2007BZT}. Other formula, such
as the Cole-Cole and the Havriliak-Negami equations, are also employed
to model the dielectric response with respect to frequency at given
temperatures. When phonon modes are close or interacting with the
relaxation modes, it becomes necessary to use coupled modes to model
the dielectric response \cite{Wang2014,Wang2016}. On the other hand,
there are also many investigations on how the dielectric response,
$\chi$, depends on the temperature, $T$, at given frequencies. In
addition to the well known Curie law for $\chi\left(T\right)$ at
high temperature, most useful equation for fitting around the dielectric
peak appears to be the square law. In particular, the formula 
\begin{align}
\frac{1}{\varepsilon\left(T\right)}= & \frac{1}{\varepsilon_{A}}+\frac{\left(T-T_{A}\right)^{^{\eta}}}{B}\label{eq:square-law}
\end{align}
was proposed to describe the permittivity at $T>T_{m}$ \cite{Smolenskii1970,Kirillov1973}.
Initially, $\eta$ was found to be 2, but later was shown to be between
1 and 2 \cite{Clarke1974,Uchino1982,Santos2001,Bokov2006a,Bokov2012}.
Here, we further the investigation in this direction and attempt to
address some important questions regarding relaxor behaviors. We will
explain why the dielectric constant has a peak value at $T_{m}$,
and what causes the asymmetry around the peak. Moreover, by constructing
a statistical model that properly describes how dipoles behave in
relaxors, we propose formula to fit experimental results, which further
illuminates the physics behind relaxation behavior.

This paper is organized as the follows. In Sec. \ref{sec:Statistical-modeling},
we introduce the statistical model. In Sec. \ref{sec:Results}, we
apply this model to both lead-free and lead-based relaxors. In Sec.
\ref{subsec:Discussion}, we discuss the implication and limitation
of this approach. Finally, in Sec. \ref{sec:Conclusion}, we present
a brief conclusion.

\section{Statistical modeling\label{sec:Statistical-modeling}}

The statistical model starts by considering a critical difference
between ferroelectric relaxors and normal ferroelectrics. One crucial
observation is that all relaxor ferroelectrics discovered so far are
inevitably disordered and inhomogeneous systems. For instance, in
BZT Zr and Ti ions are distributed randomly, so are the Mg and Nb
ions in PMN, when the samples are treated macroscopically. In addition,
PMN possesses the electric field arising from heterovalent Mg and
Nb ions, which affects dipole distribution. It is important to further
note that well known relaxors can become non-relaxor if their ions
are perfectly ordered \cite{Setter1980,Bokov1984,WangJAD}.

\subsection{Individual dipoles}

The randomness of ions and the ensuing lack of long-range correlation
has the important consequence that phonon modes may not be the best
description to understand relaxor. This fact is evidenced by the effective
Hamiltonian that describes the BZT relaxor \cite{Akbarzadeh2012,Sherrington2013a,Sherrington2014}
\begin{align}
E= & \sum_{i}\left(\kappa_{i}\left|u_{i}\right|^{2}+\lambda_{i}\left|u_{i}\right|^{4}\right)+\dots,\label{eq:effHBZT}
\end{align}
where $i$ labels the sites occupied randomly by Zr or Ti, and $\kappa_{i}$
($\lambda_{i}$) are the second (fourth) order coefficients in the
Taylor expansion of energy with respect to $u_{i}$, which is the
local dipole on site $i$. For a homogeneous system, where $k_{i}$
and $\lambda_{i}$ are constants, we can usually first consider the
harmonic term and construct phonon modes, which are then used to describe
the system, especially in low temperature when the system condense
to particular phonon modes \cite{CondenedPhononMode}. In contrast,
with the loss of periodicity in relaxors, this approach is no longer
profitable. One can insist on using averaged atoms (e.g., replacing
Zr and Ti atoms with their average in BZT) to retain the use of phonon
modes. However, it is then necessary to consider defect-pinned intrinsic
localized modes \cite{PinnedPhonon} and phonon localization \cite{PhononLocalization}. 

The inhomogeneity also has important consequences on ferroelectric
phase transitions. In the typical ferroelectric material BaTiO$_{3}$,
we may ascribe the temperature-driven phase transition to the condense
of phonons to a particular phonon mode \cite{CondenedPhononMode}.
At high temperature, many phonons modes are occupied (occupancy obeying
the Bose-Einstein distribution); at low temperatures, due to mode
softening, certain mode (often corresponding to the well-known \textcolor{black}{soft
mode \cite{Zhong1994,StruPhaseTran,Zhong1995}} in perovskites) has
essentially zero energy, which dominates the system and induce phase
transitions. Unlike BaTiO$_{3}$, there is no global phase transition
due to the existence of PNRs and/or random electric fields, which
eliminates the mode softening phenomenon and renders a global dipole
pattern difficult to achieve \cite{Akbarzadeh2012,PMNeffHami2015}.
In addition, relaxors exhibit strange phonon behavior, such as the
``waterfall'' effect \cite{Gehring2000,Hlinka2003,PMNWaterFall}
and the localization\cite{PhononLocalization}, further showing their
difference from normal ferroelectrics. In this work, we focus on individual
dipoles and statistically model their dielectric response. This change
of view point implies that phonon modes are less important in our
analysis. We will show in the following that such change leads to
fruitful results, and better understanding of relaxors.

\subsection{Statistics of individual dipoles}

Individual dipoles can be categorized into different groups based
on their dynamics, and each group shall have different contribution
to susceptibility. We proceed simplify the interaction between dipoles,
assuming that the interaction effectively introduce a potential well
of \emph{average} depth, $E_{b}$. We may relate $E_{b}$ to the size
of PNRs arising from the clustering of same-type ions \cite{Akbarzadeh2012}
and/or random electric field caused by heterovalent ions \cite{PMNeffHami2015,Kleemann2014,PMNWaterFall}. 

Since the kinetic energy of individual dipoles obeys the Maxwell-Boltzmann
distribution,  at temperature $T$, the number of dipoles with kinetic
energy $E_{\textrm{kin}}$ is given by
\begin{align}
f\left(E_{\textrm{kin}}\right)= & 2N\sqrt{\frac{E_{\textrm{kin}}}{\pi}}\left(\frac{1}{k_{B}T}\right)^{3/2}\exp\left(-\frac{E_{\textrm{kin}}}{k_{B}T}\right),\label{eq:kinetic-energy}
\end{align}
where $k_{B}$ is the Boltzmann constant, $N$ is the total number
of dipoles, and $f\left(E_{\textrm{kin}}\right)dE_{\textrm{kin}}$
is the number of dipoles having a kinetic energy between $E_{\textrm{kin}}$
and $E_{\textrm{kin}}+dE_{\textrm{kin}}$. With this distribution
function, we can calculate the number of dipoles with kinetic energy
that exceeds the potential well $E_{b}$, which is given by
\begin{align}
N_{1}\left(E_{b},T\right) & =\int_{E_{b}}^{\infty}dE_{\textrm{kin}}f\left(E_{\textrm{kin}}\right)\nonumber \\
= & N\sqrt{\frac{4}{\pi}}\sqrt{\frac{E_{b}}{k_{B}T}}\exp\left(-\frac{E_{b}}{k_{B}T}\right)+N\text{erfc}\left(\sqrt{\frac{E_{b}}{k_{B}T}}\right),\label{eq:particles-above-well}
\end{align}
where erfc is the complementary error function. The number of dipoles
confined to the potential well is then given by
\begin{align}
N_{2}\left(E_{b},T\right)= & N-N_{1}\left(E_{b},T\right).\label{eq:particles-inside-well}
\end{align}

The next step is to treat the two sets of dipoles ($N_{1}$ versus
$N_{2}$) separately, assigning different susceptibility to them.
The total susceptibility is then given by 
\begin{align}
\chi(T,\nu)= & \chi_{1}\left(T,\nu\right)P_{1}\left(E_{b},T\right)+\chi_{2}\left(T,\nu\right)P_{2}\left(E_{b},T\right),\label{eq:total-susceptibility}
\end{align}
where $\chi_{1}\left(T,\nu\right)$ and $\chi_{2}\left(T,\nu\right)$
describes the dielectric responses of each dipole group, whose form
will be specified later. We also define 
\begin{align*}
P_{1}\left(E_{b},T\right)\equiv & \frac{1}{N}N_{1}\left(E_{b},T\right),\\
P_{2}\left(E_{b},T\right)\equiv & \frac{1}{N}N_{2}\left(E_{b},T\right),
\end{align*}
to normalize the dipoles to unit volume. Equation (\ref{eq:total-susceptibility})
is the centerpiece of this work and will be demonstrated to be useful
for investigation of various relaxors.

\section{Results \label{sec:Results}}

We now apply Eq. (\ref{eq:total-susceptibility}) to fit various $\chi$
versus $T$ obtained experimentally or numerically. The relaxors shown
here include both lead-basd (e.g., PMN) and lead-free relaxors (e.g.,
BZT).

\subsection{Susceptibility of BZT \label{sec:lead-free-relaxor}}

For the static susceptibility of lead-free relaxor BZT\cite{BZT-susceptibility,Prosandeev dielectric-susceptibility,Petzelt2014,Wang2016},
we assume (i) dipoles with kinetic energy that overcomes potential
well can be treated as free dipoles, subject only to thermal excitation;
(ii) dipoles inside the potential well only contribute a constant
susceptibility, $\chi_{2}$. The total susceptibility is thus given
by
\begin{align}
\chi(T)= & \chi_{1}\mathcal{L}\left(\frac{\theta}{T}\right)P_{1}\left(E_{b},T\right)+\chi_{2}P_{2}\left(E_{b},T\right)\label{eq:BZT-susceptibility}
\end{align}
where $\mathcal{L}\left(x\right)=\coth\left(x\right)-1/x$ is the
Langevin function, known for depicting orientational polarization
under thermal fluctuations \cite{Kasap2006,CurieLaw}. $E_{b}$, $\chi_{1}$,
$\chi_{2}$ and $\theta$ are constants, which will be determined
by fitting experimental or numerical data. It can be inferred from
equation \ref{eq:BZT-susceptibility} that $\chi_{1}$ is the susceptibility
of the material at zero Kevin, $\chi_{1}\mathcal{L}\left(\frac{\theta}{T}\right)$
is essentially the Curie law at high temperature, and $\theta$ is
proportional to the magnitude of the low-frequency electric field
used in experimental measurements.

\begin{figure}[h]
\noindent \begin{centering}
\includegraphics[width=8cm]{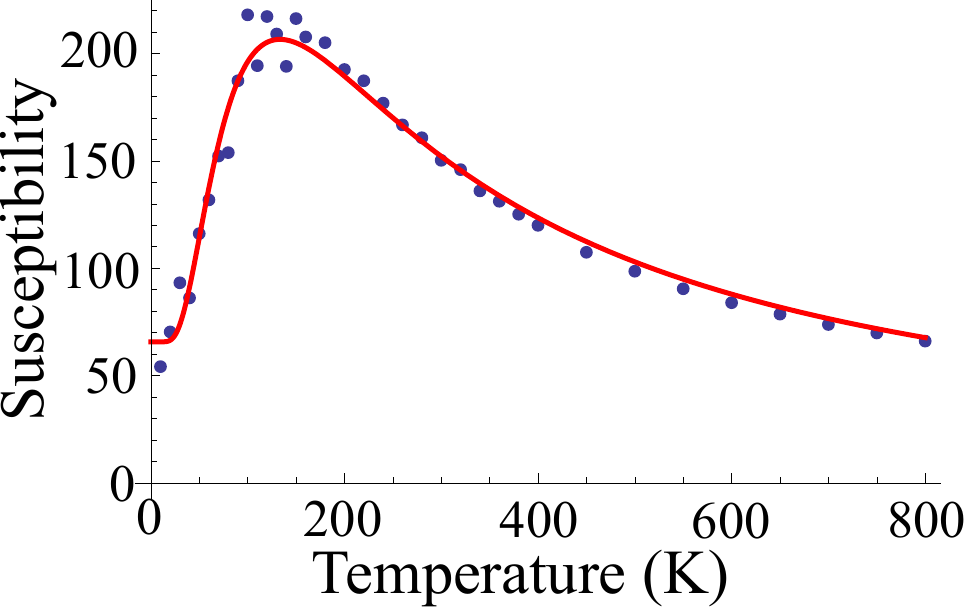}
\par\end{centering}
\caption{Fitting the static susceptibility of Ba(Zr$_{0.5}$,Ti$_{0.5}$)O$_{3}$
obtained from Monte-Carlo simulation using Eq. (\ref{eq:BZT-susceptibility}).
The blue dots are from Monte-Carlo simulation \cite{Akbarzadeh2012}
and the red solid line is the fitting curve using Eq. (\ref{eq:BZT-susceptibility}).
\label{fig:Fitting-susceptibility-of}}
\end{figure}

We first test Eq. (\ref{eq:BZT-susceptibility}) against the static
susceptibility versus temperature obtained with Monte-Carlo (MC) simulation
in a previous work \cite{Akbarzadeh2012}. \textcolor{black}{Figure}\textcolor{lime}{{}
}\ref{fig:Fitting-susceptibility-of} shows the overall fitting is
good enough to reproduce results from MC simulations with parameters
shown in and Tab. \ref{tab:parameters-BZT-static}. The closeness
of $E_{b}$ and $T_{m}$ indicates the average depth of potential
wells plays a dominant role in determining $T_{m}$. Close examination
of Fig. \ref{fig:Fitting-susceptibility-of} also indicates the fitting
at the lowest temperature ($\lesssim25$ K) is not as good as the
rest. To address this issue, we tried adding a Gaussian distribution
to $E_{b}$ and remedied the minor problem. However, the resulting
equation is quite complicated and deviates from our original goal
of proposing simple analytical formula to fit susceptibility. Therefore
this additional step is not adopted here.
\begin{table}
\noindent \begin{centering}
\begin{tabular}{|c|c|c|c|c|}
\hline 
 & $\chi_{1}$ & $\theta$ (K) & $\chi_{2}$ & $E_{b}$ (K)\tabularnewline
\hline 
\hline 
Values & 741.6 & 220.5  & 64.7 & 159.1\tabularnewline
\hline 
\end{tabular}
\par\end{centering}
\caption{Fitting parameters for the Ba(Zr$_{0.5}$,Ti$_{0.5}$)O$_{3}$ static
susceptibility. \label{tab:parameters-BZT-static}}
\end{table}

\begin{figure}[h]
\noindent \begin{centering}
\includegraphics[width=6cm]{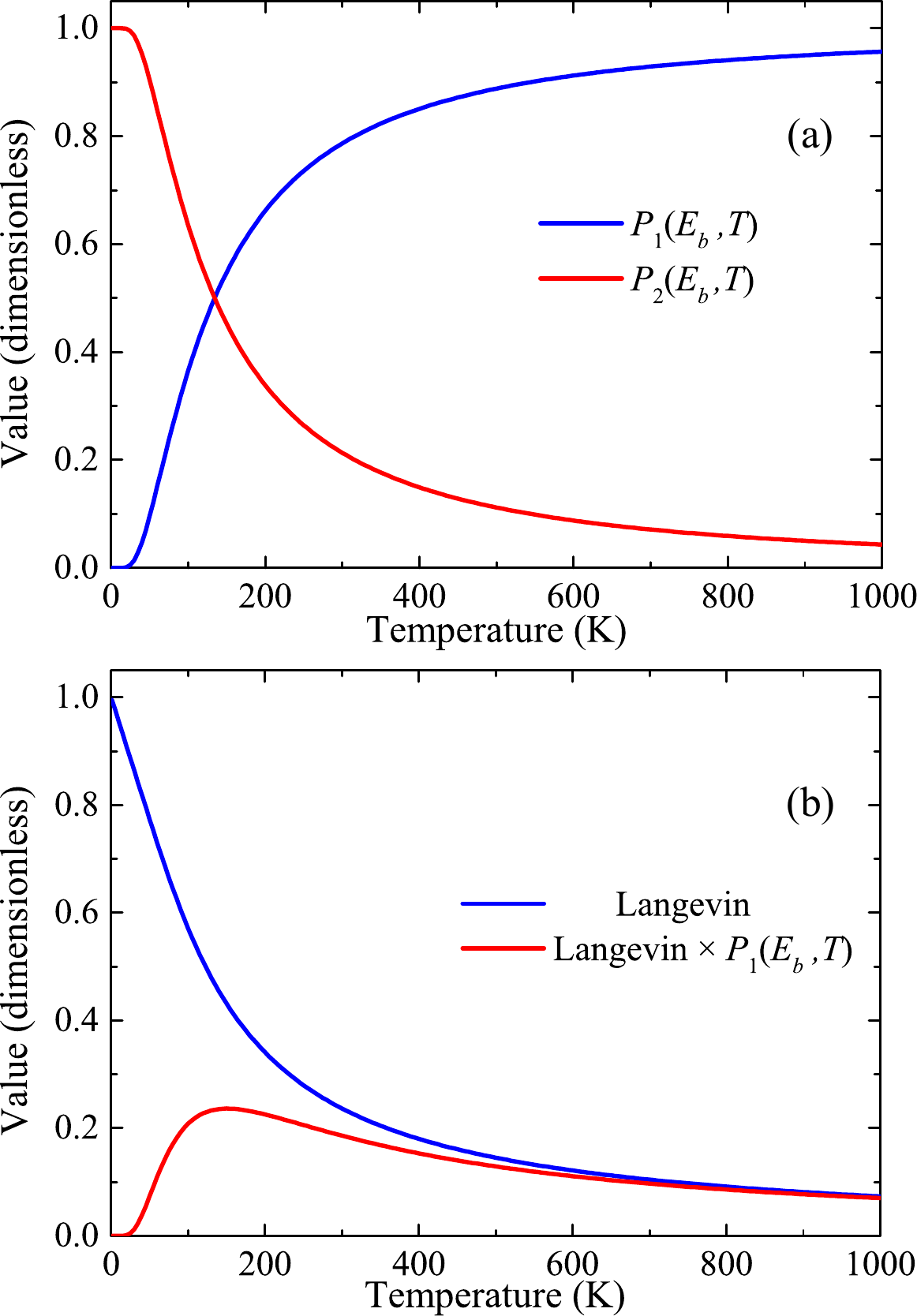}
\par\end{centering}
\caption{Maxwell-Boltzmann distribution {[}Panel (a){]} and the Langevin function,
$\mathcal{L}$, {[}Panel (b){]} versus temperature. Parameters from
Tab. \ref{tab:parameters-BZT-static} are used in plot each function.
\label{fig:components}}
\end{figure}
 In order to have a good understanding of BZT's susceptibility, we
show each component of Eq. (\ref{eq:BZT-susceptibility}) in Fig.
\ref{fig:components}. Figure \ref{fig:components} (a) shows that
$P_{1}\left(E_{b},T\right)$ and $P_{2}\left(E_{b},T\right)$ have
opposite trends as temperature increases. The number of dipoles that
can overcome the potential confinement ($P_{1}$) steadily increases
with temperature, while the number of dipoles inside ($P_{2}$) continuously
becomes smaller. Figure \ref{fig:components} (b) shows that the Lagevin
function is normalized at $T=0$, and decreases with temperature.
Such a feature describes the ability of the free dipoles to respond
to an external \emph{dc} electric field. Moreover, Fig. \ref{fig:components}
(b) also shows the product of the Langevin function and $P_{1}$,
which already exhibits some resemblance to BZT's susceptibility {[}Fig.
(\ref{fig:Fitting-susceptibility-of}){]}.

\begin{figure}[h]
\noindent \begin{centering}
\includegraphics[width=8cm]{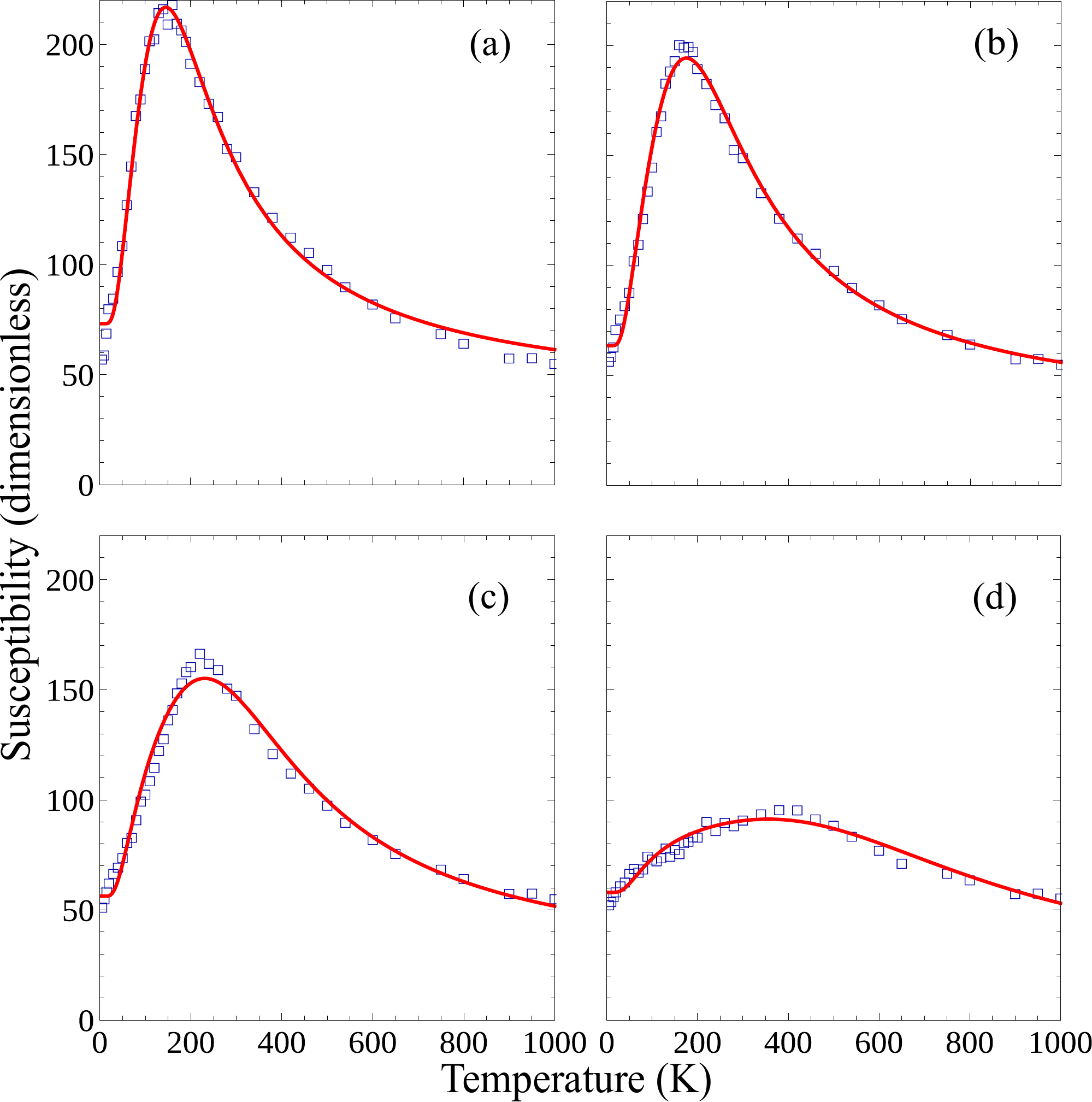}
\par\end{centering}
\caption{Fitting the susceptibility of Ba(Zr$_{0.5}$,Ti$_{0.5}$)O$_{3}$
at $f=1,10,100,100$ GHz, obtained from molecular dynamics simulation
using Eq. (\ref{eq:BZT-frequency}).\label{fig:Fitting-BZT-frequency}}
\end{figure}
 Having examined the static susceptibility, we now move to the frequency-dependent
dielectric response, which is often taken as a characteristic property
of relaxors \cite{Kleeman2014,Sommer}. We propose another equation
to fit the susceptibility versus temperature:

\begin{align}
\chi\left(T\right)= & \frac{\chi_{1}}{1+b\exp\left(-\theta/T\right)}P\left(E_{b},T\right)+\chi_{2}\left[1-P\left(E_{b},T\right)\right],\label{eq:BZT-frequency}
\end{align}
where $\chi_{1}$ , $\chi_{2}$, $b$ and $\theta$ are constants
at a given frequency (but may change when the frequency changes).
The dielectric contribution from dipoles with kinetic energy higher
than the potential well is given by
\begin{align}
w_{1}\left(T\right)= & \frac{1}{1+b\exp\left(-\theta/T\right)}.\label{eq:h1T}
\end{align}
which, similar to the Langevin function, monotonically decreases with
temperature $T$. The choice of this function reflects two considerations:
(i) at very low temperature ($T$ close to 0), such dipoles shall
follow the probing $ac$ electric field closely, leading $w_{1}\left(T\right)$
to its maximum; and (ii) at higher temperature, thermal motions of
these dipoles hamper their ability to follow the \emph{ac} electric
field, leading to smaller $w_{1}\left(T\right)$. We will discuss
this equation further in Sec. \ref{subsec:Discussion}. With one more
parameter ($b$), this function may be taken as an extension to the
Langevin function.

\begin{table}[h]
\noindent \centering{}%
\begin{tabular}{|c|c|c|c|c|}
\hline 
 & 1 GHz & 10 GHz & 100 GHz & 1000 GHz\tabularnewline
\hline 
\hline 
$\theta$ (K) & 579.6 & 762.6 & 1128.4 & 2158\tabularnewline
\hline 
$\chi_{1}$  & 406.5 & 312.6 & 209.3 & 99.9\tabularnewline
\hline 
$\chi_{2}$ & 73.1 & 63.5 & 56.2 & 57.9\tabularnewline
\hline 
$b$ & 10.2 & 9.9 & 9.5 & 7.7\tabularnewline
\hline 
\end{tabular} \caption{Fitting parameters of numerically simulated Ba(Zr$_{0.5}$,Ti$_{0.5}$)O$_{3}$
susceptibility at various frequency \cite{Wang2016} using Eq. (\ref{eq:BZT-frequency}).
\label{tab:Fitting-parameters-of-BZT}}
\end{table}
\begin{figure}[h]
\noindent \begin{centering}
\includegraphics[width=8cm]{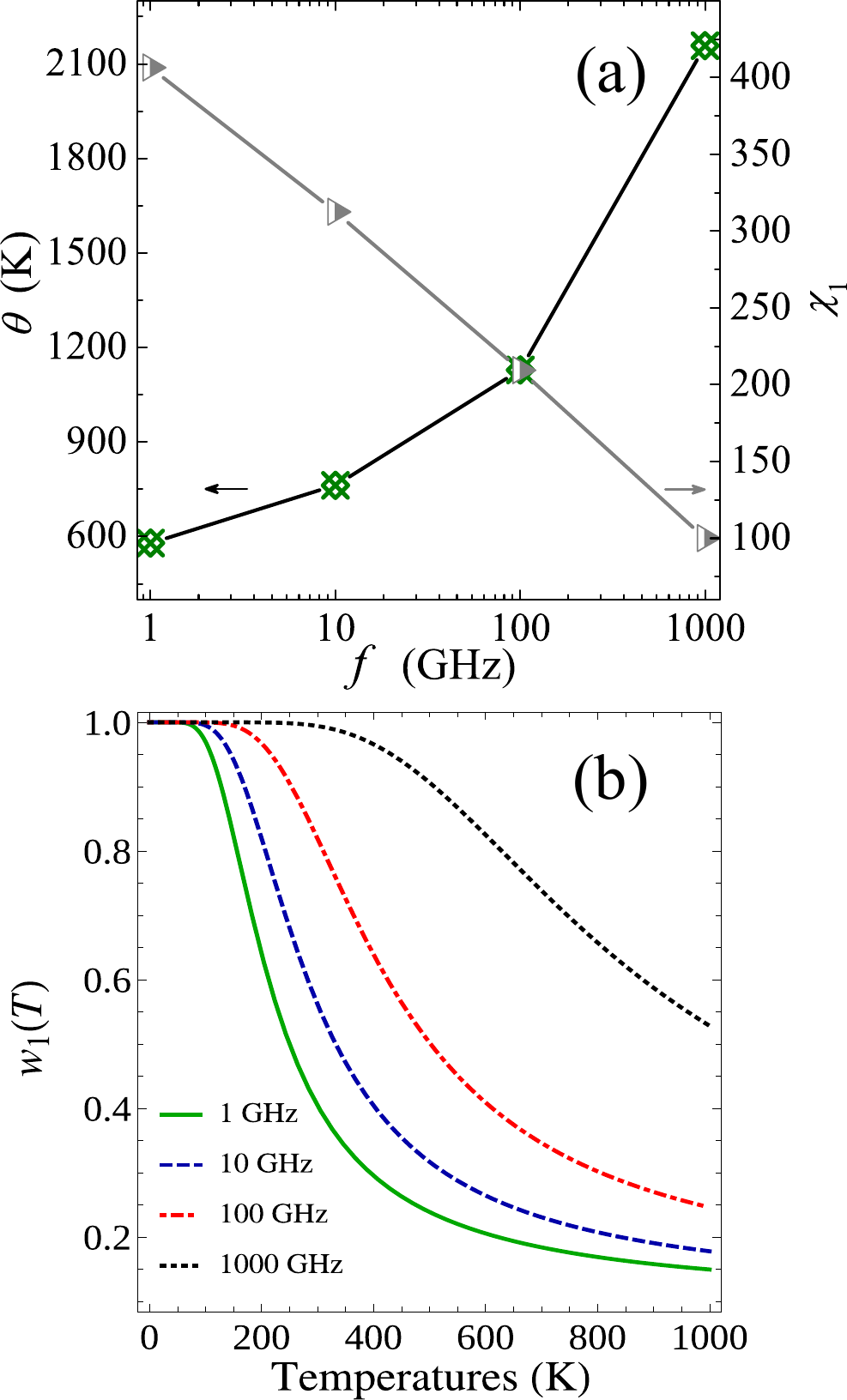}
\par\end{centering}
\caption{Analysis of fitting parameters versus probing frequency. (a) \textcolor{black}{$\theta$
vs $\log(f)$ and} $\chi_{1}$vs $\log\left(f\right)$; (b) $w_{1}\left(T\right)$.
\label{fig:w1T}}
\end{figure}
 We use Eq. (\ref{eq:BZT-frequency}) to fit BZT's susceptibility
versus temperature at different frequency and show the results in
Fig. \ref{fig:Fitting-BZT-frequency}. Since $E_{b}$ is a material
parameter, we use the same value ($E_{b}=159.1$ K) obtained by fitting
the static susceptibility (cf. Fig. \ref{fig:Fitting-susceptibility-of}).
In Fig. \ref{fig:Fitting-BZT-frequency}, the numerical results are
obtained from molecular dynamics simulations reported in Ref. \cite{Wang2016}.
As the figure shows, satisfactory fittings are achieved for frequencies
between 1 and 1000 GHz. Table \ref{tab:Fitting-parameters-of-BZT}
shows all the parameters. Among them, $\theta_{1}$ and $\chi_{1}$
have substantial changes over the specified frequency range as shown
in Fig \ref{fig:w1T}(a). Figures \ref{fig:w1T} (a) shows that $\theta$
depends on $\log\left(f\right)$ quadratically while $\chi_{1}$ linearly
depends on it, and Fig. \ref{fig:w1T}(b) shows $w_{1}\left(T\right)$.
At low frequency ($\lesssim10$ GHz), $w_{1}\left(T\right)$ resembles
the Fermi-Dirac function, that is, below$\sim250$ K, its value is
close to one but becomes close zero for $T$ above $\sim250$ K. At
a higher frequency (e.g., 1000 GHz), however, this function strongly
deviates from the Fermi-Dirac function, with a long tail extending
to high temperature.

\begin{figure}[h]
\noindent \begin{centering}
\includegraphics[width=6cm]{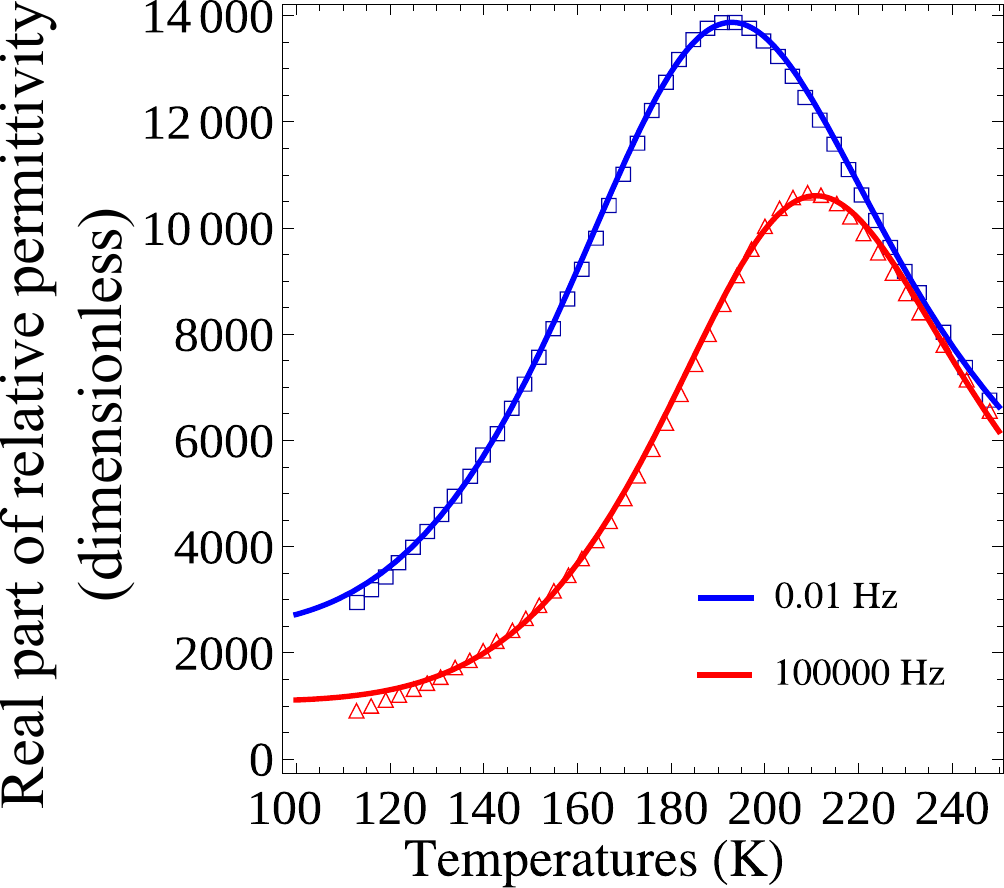}
\par\end{centering}
\caption{Experimental susceptibility of Ba(Ti$_{0.675}$Zr$_{0.325}$)O$_{3}$
ceramics versus temperature fitted with Eq. (\ref{eq:BZT-frequency}).
\label{fig:real-bzt}}
\end{figure}
To further verify the suitability of this equation for experimental
data, we also fit the result shown in Fig. 1 of Ref. \cite{Bokov2007BZT},
where Ba(Ti$_{0.675}$Zr$_{0.325}$)O$_{3}$ ceramics is measured
at $10^{-2}$ and $10^{5}$ Hz. Figure \ref{fig:real-bzt} shows that
satisfactory fittings are achieved.

\subsection{Pb-based relaxors \label{subsec:Lead-based-relaxors}}

 Unlike the lead-free BZT, which possesses PNRs that separate dipole
clusters, lead-based ferroelectrics \cite{Kirillov1973,Gridnev2004,PMNeffHami2015,PNR-PMN}
have the Pb-driven dipoles across the system \cite{Yk}, which cause
phase transitions in systems such as PbTiO$_{3}$ and Pb(Zr,Ti)O$_{3}$
\cite{PZTPhase,PTOeffHam,JiangPTOFilm-1}. Due to heterovalent ions
inside, typical lead-based relaxors are subject to random electric
fields, which distort the orientation of dipoles. While the precise
consequence of the random field is not all clear \cite{Sherrington2014,Kleemann2015},
such distracting effect on dipoles appears to lead to a strongly modified
phase transition with diffused and smeared peak, in sharp contrast
to that of normal ferroelectrics \cite{Pirc2007,PMNeffHami2015}. 

To model such a system and account for the moderate phase transition,
we need a function that properly describes the dielectric constant
versus temperature. Here, we propose to use the slightly modified
well known quadratic relation $\frac{\varepsilon_{A}}{\varepsilon^{\prime}}-1=\frac{\left(T-T_{A}\right)^{2}}{2\delta^{2}}$
proposed by Bokov \emph{et al} \cite{Bokov2006a} to relate relaxor's
permittivity to temperature \cite{Smolenskii1970,Kirillov1973} for
dipoles above the average potential well (also see Eq. (\ref{eq:square-law})
). This equation can be rearranged to give the following expression
\begin{align}
w_{2}\left(T\right)= & \frac{1}{1+\left|\frac{T-T_{O}}{\theta}\right|^{\gamma}},\label{eq:w2t}
\end{align}
where $\gamma$ is a critical exponent, $T_{o}$ is associated with
the peak position of the moderate phase transition, $\theta$ and
$\gamma$ are parameters describing the peak. We note that such choice
of $w_{2}\left(T\right)$ also agree with the analysis recently given
by Uchino \cite{Uchino2014}. Combining Eqs. (\ref{eq:total-susceptibility})
and (\ref{eq:w2t}), we obtain the following equation to fit lead-based
relaxors,
\begin{align}
\chi\left(T\right)= & \frac{\chi_{1}}{1+\left|\frac{T-T_{O}}{\theta}\right|^{\gamma}}P\left(E_{b},T\right)+\chi_{2}\left[1-P\left(E_{b},T\right)\right],\label{eq:PZN}
\end{align}
where $E_{b}$, $\chi_{1}$, $\chi_{2}$, $T_{o}$, $\theta$, and
$\gamma$ are fitting parameters. The meaning of $\chi_{1}$, $\chi_{2}$,
and $E_{b}$ are the same as discussed in Sec. \ref{sec:lead-free-relaxor}. 

\begin{figure}[h]
\noindent \begin{centering}
\includegraphics[width=6cm]{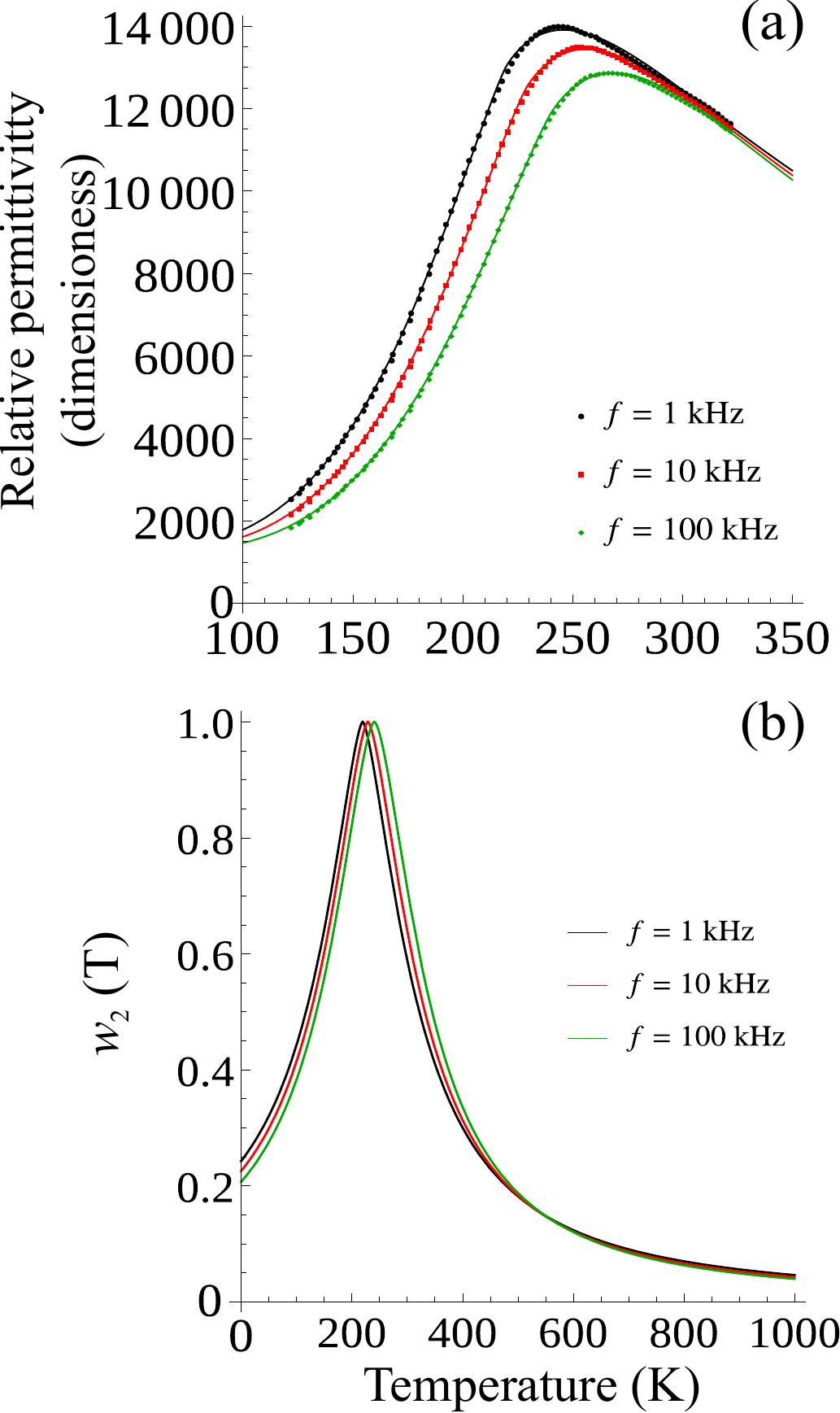}
\par\end{centering}
\caption{Fitting the relative permittivity of PZN-0.13PT at $f=1,10,100$ kHz
using Eq. (\ref{eq:PZN}). Note the abnormal decrease in the range
below $\sim250$ K. The second panel is $w_{2}\left(T\right)$. \label{fig:PZN-relaxor}}
\end{figure}
 To verify that Eq. (\ref{eq:PZN}) indeed works, we experimentally
obtained the permittivity of Pb(Zn$_{1/3}$Nb$_{2/3}$)$_{0.87}$Ti$_{0.13}$O$_{3}$
(PZN-0.13PT) versus temperature at frequencies $f=1,10,100$ kHz.
As Fig. \ref{fig:PZN-relaxor}(a) shows, all three fittings are satisfactory.
With Eq. (\ref{eq:PZN}) and the fittings, we are able to single out
$w_{2}\left(T\right)$, which shows a slight increase of the peak
temperature ($T_{o}$, around $220$ K) with frequency {[}Fig. \ref{fig:PZN-relaxor}(b){]}.
We note that in this fitting there is no need to have two $\gamma$
values above and below $T_{o}$ \cite{Uchino2014}. The asymmetric
peak shown in Fig. \ref{fig:PZN-relaxor}(a) is naturally caused by
the Maxwell-Boltzmann distribution function.

\begin{table}[h]
\noindent \centering{}%
\begin{tabular}{|c|c|c|c|}
\hline 
 & 1 kHz & 10 kHz & 100 kHz\tabularnewline
\hline 
\hline 
$\gamma$ & 2.02 & 1.82 & 1.63\tabularnewline
\hline 
$T_{O}$ (K) & 219.6 & 229.0 & 240.6\tabularnewline
\hline 
$E_{b}$ (K) & 22.3 & 22.9 & 23.5\tabularnewline
\hline 
$\chi_{1}$ \footnote{Here, $\chi_{1}$ and $\chi_{2}$ shall be understood as relative
permittivity, not susceptibility. } & 56601 & 55781 & 53529\tabularnewline
\hline 
$\chi_{2}$ & 1320.9 & 1284.2 & 1238.0\tabularnewline
\hline 
$\theta$ (K) & 102.4 & 103.1 & 104.5\tabularnewline
\hline 
\end{tabular}\caption{Fitting parameters of PZN-0.13PT's permittivity measured at different
frequencies. As the table demonstrates, $\gamma$ and $T_{C}$ are
the most important variable that changes a lot with frequency. \label{tab:Fitting-parameters-of-PZN}}
\end{table}
 Table \ref{tab:Fitting-parameters-of-PZN} summarizes fitting parameters
of the permittivity measured at various frequencies. Among all the
parameters, the critical component $\gamma$ changes most (19.3\%
from 1 kHz to 100 kHz), and decreases with increasing frequency; similarly,
$T_{O}$ also changes by 9.5\%. On the other hand, $E_{b}$, $\chi_{1}$,
$\chi_{2}$, $\theta$ are relatively constant, which are independent
of the frequency, and may be taken as material parameters. Such results
hints towards the following formula that describes the dependence
of PZN-0.13PT on both temperature and the probing frequency
\begin{align}
\chi\left(T,\nu\right)= & \frac{\chi_{1}}{1+\left|\frac{T-T_{O}\left(\nu\right)}{\theta}\right|^{\gamma\left(\nu\right)}}P\left(E_{b},T\right)+\chi_{2}\left[1-P\left(E_{b},T\right)\right],\label{eq:PZN-1}
\end{align}
 where the two functions $T_{O}\left(\nu\right)$ and $\gamma\left(\nu\right)$
are frequency dependent while other parameters are constants for a
given material. It is also worth noting that for PZN-0.13PT $\chi_{2}\ll\chi_{1}$,
which indicates that dipoles with kinetic energy above the potential
well play a more important role, in contrast to the case of BZT (see
Tab. \ref{tab:Fitting-parameters-of-BZT}).

\begin{figure}[h]
\noindent \begin{centering}
\includegraphics[width=6cm]{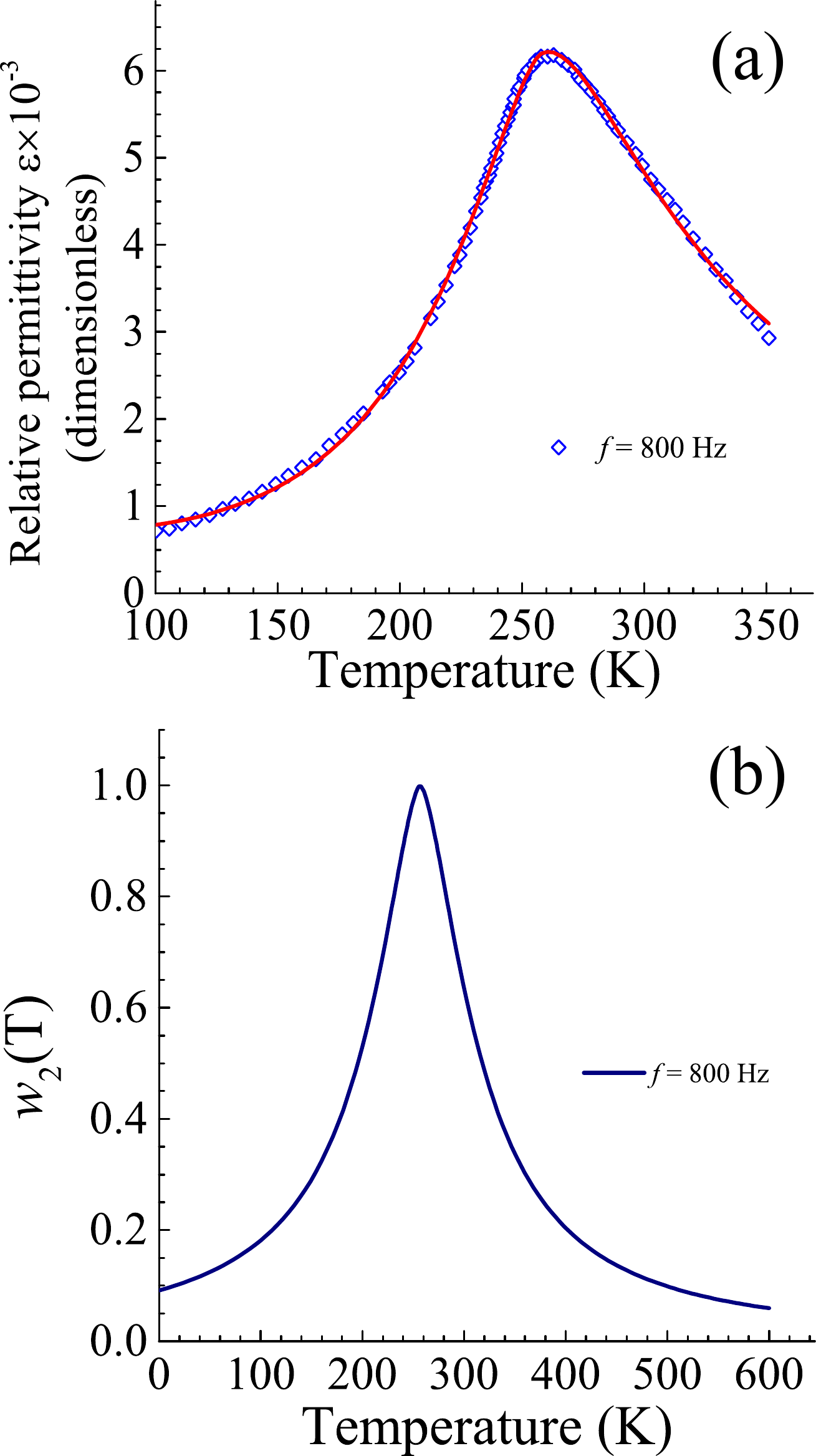}
\par\end{centering}
\caption{Fitting the relative permittivity of Pb(Mg$_{1/3}$Nb$_{2/3}$)O$_{3}$-0.05Pb(Zr$_{0.53}$Ti$_{0.47}$)O$_{3}$
measured at $800$ Hz using Eq. (\ref{eq:PZN}). The second panel
is $w_{2}\left(T\right)$. \label{fig:PMN-PZT}}
\end{figure}
To further verify the proposed formula, we also fit the permittivity
versus temperature of another lead-based relaxor, Pb(Mg$_{1/3}$Nb$_{2/3}$)O$_{3}$-0.05Pb(Zr$_{0.53}$Ti$_{0.47}$)O$_{3}$
\cite{Gridnev2004}. It can be seen from Fig. \ref{fig:PMN-PZT}(a)
that the overall fitting is satisfactory. Figure \ref{fig:PMN-PZT}(b)
shows $w_{2}\left(T\right)$ with fitting parameters $\gamma=1.66$
and $T_{O}=256.7$ K. Similar to the PZN-0.13PT case, the results
here also shows $\chi_{2}\ll\chi_{1}$.

\section{Discussion\label{subsec:Discussion}}

In the statistical model we divide the dipoles inside ferroelectrics
relaxors into two groups, one group being confined in potential wells,
while the other having can overcome the potential confinement and
exhibiting a more vibrant dynamics. It has been demonstrated that
the Maxwell-Boltzmann distribution plays a significant role in determining
the profile of $\chi\left(T\right)$. To address a particular type
of relaxor, one may only need to adjust the dielectric response function
associated with each group of dipoles, while keeping the rest unchanged.
In this section, we discuss a few issues of this approach and its
limitations.

\subsection{Characteristic temperature $T_{m}$}

The present analysis helps us to understand why the susceptibility
of a relaxor reaches its peak value at some temperature, $T_{m}$.
For BZT, the function $\chi_{1}\mathcal{L}\left(\frac{\theta}{T}\right)$
{[}see Eq. (\ref{eq:BZT-susceptibility}) and Fig. \ref{fig:components}{]}
or $\chi_{1}/\left[1+b\exp\left(-\theta/T\right)\right]$ {[}see Eq.
(\ref{eq:BZT-frequency}){]} describes the contribution to susceptibility
from dipoles with kinetic energy higher than $E_{b}$. These two functions
are both monotonically decreasing with $T$, reflecting the fact that
thermal motions prevent dipoles from aligning with the applied electric
field. On the other hand, the number of dipoles above potential wells
increases with $T$ as governed by the function $P\left(E_{b},T\right)$
(see Fig. \ref{fig:components}). The combined effects of the two
factors give rise to the susceptibility peak at $T_{m}$. However,
the situation for lead-based relaxors is different. The function $\chi_{1}/\left(1+\left|\frac{T-T_{O}}{\theta}\right|^{\gamma}\right)$
{[}see Eq. (\ref{eq:PZN}){]}, which largely determines the value
of $T_{m}$, manifests the vestige of a true phase transition in normal
ferroelectrics, which is torn down by random electric fields and/or
PNRs in relaxors.

\subsection{Rationale for Eq. (\ref{eq:h1T})}

\begin{figure}[h]
\centering{}\includegraphics[clip,width=8cm]{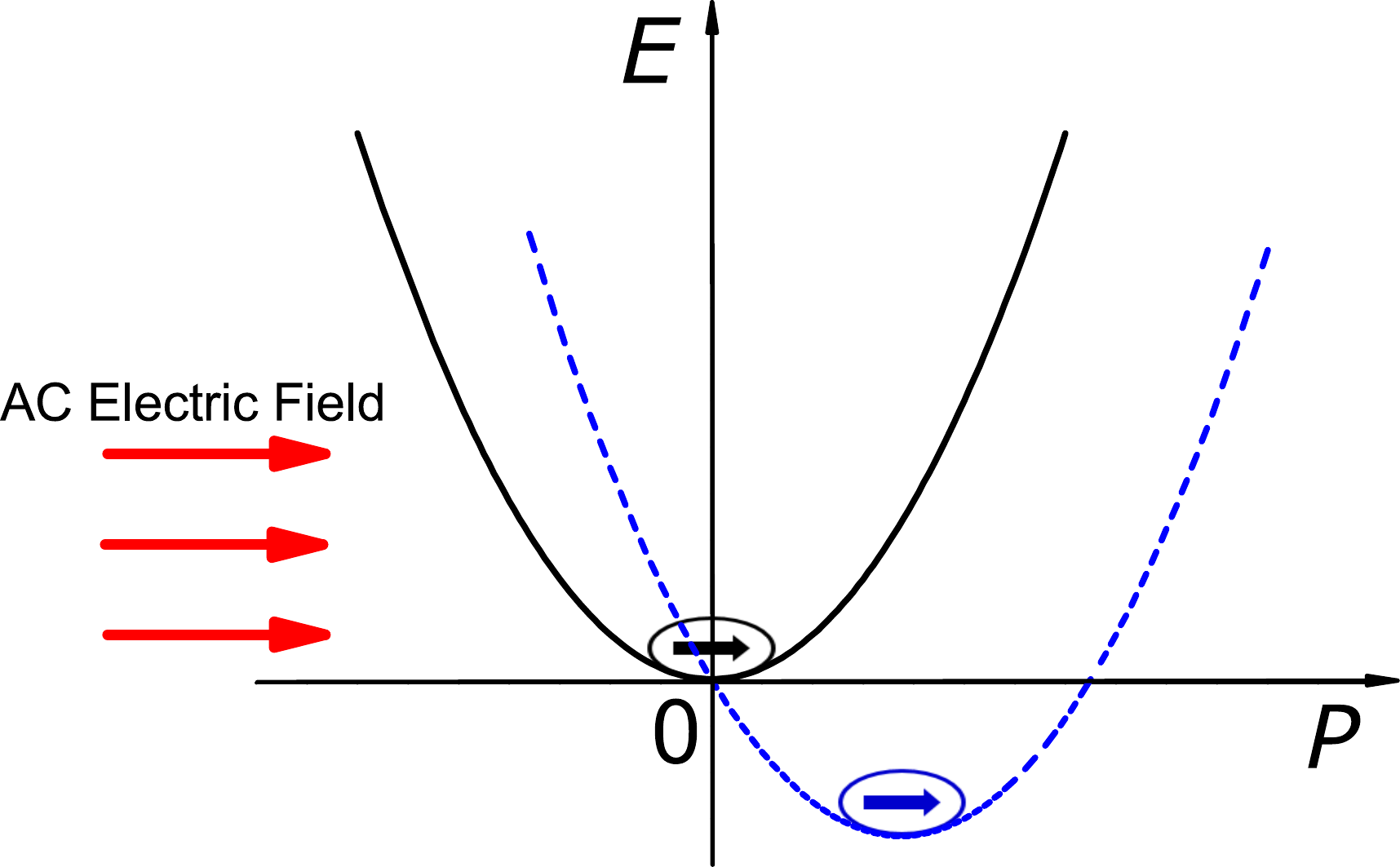}\caption{The energy well for dipoles is shifted and lowered when an electric
field is applied .\label{fig:falling}}
\end{figure}
 For lead-free BZT, we propose Eq. \ref{eq:h1T} to describe the susceptibility
of dipoles with kinetic energy higher than $E_{b}$. This choice follows
the Debye relaxation, i.e., $\chi\sim1/\left(1+\omega^{2}\tau^{2}\right)$
\cite{Kasap2006}, where $\omega$ is a constant (the probing frequency),
and $\tau$ is temperature-dependent relaxation time. For a thermally
activated process, the relation between $\tau$ and $T$ is often
specified by the Arrhenius law, i.e., $1/\tau=A\exp\left(-E_{a}/T\right)$,
where $E_{a}>0$ is the activation energy \cite{PinnedPhonon,negativeActivation,Pirc2007}.
In this case, the susceptibility will be $\chi\sim1/\left[1+A^{2}\omega^{2}\exp\left(\frac{2E_{a}}{T}\right)\right]$,
which is discussed by Jonscher \cite{Jonscher1981}. However, the
dynamic process considered here describes dipoles falling to a state
of lower energy, which temporarily created by the probing electric
field (see. Fig. \ref{fig:falling}). Therefore the activation energy
in this process shall be \emph{negative}, i.e., $\chi\sim1/\left[1+A^{2}\omega^{2}\exp\left(-\frac{2\left|E_{a}\right|}{T}\right)\right]$,
which is the form adopted in Eq. (\ref{eq:h1T}).

We note that negative activation energy is known in some chemical
reactions \cite{negativeActivation}. Negative activation energy appears
here because when an \emph{ac} electric field perturbs dipoles and
tilts the relative energy of potential wells, dipoles outside potential
wells will move towards the temporary potential minimum. However,
the drifting to the potential minimum is hindered by thermal fluctuations
of such dipoles. In fact, higher temperature (corresponding to larger
kinetic energy) results in a slower relaxation to the energy minimum
(corresponding to larger $\tau$), leading to negative activation
energy. We also note that since the applied \emph{ac} electric field
is responsible for shifting the energy minimum and causing dipoles
to drift, the change of its frequency may well alter the negative
activation energy, explaining why $\theta$ in Eq. (\ref{eq:BZT-frequency})
is dependent on the probing frequency. Similar arguments explains
why $T_{O}$ also depends on the probing frequency. 

In the above analysis, we focus on dipoles with kinetic energy higher
than $E_{b}$. These dipoles are able to drift from one energy minimum
to another when an \emph{ac} electric field perturbs the system. 
It has been proposed that the drifting/hopping of dipoles from one
potential well to another causes relaxations . However, without distinguishing
dipoles inside and outside the potential well, such proposal seems
to have a tendency of confusing the wait time before hopping, $t$
(which reflects the distribution of dipoles at a given temperature)
with the relaxation time, $\tau$ (which reflects how fast dipoles
drift to the transient energy minimum and relates to the loss peak
frequency in the Debye function), leading to some difficulties. 

\subsection{Limitations}

In previous studies \cite{Wang2014,Wang2016,PMNeffHami2015}, \emph{ab
initio} calculation was used, which prescribes all important interactions
between dipoles and other degrees of freedom in relaxors, and MC or
MD was used to numerically work out the consequences. In the present
work we do not start from \emph{ab initio} calculation, instead, employs
statistical and phenomenological arguments. Having shown results and
insights obtained with this approach, we now discuss possible limitations
to the present approach with respect to treatment of details, accuracy,
and prediction power.

First, the proposed equations for lead-free {[}Eq. (\ref{eq:BZT-frequency}){]}
and lead-based relaxors {[}Eq. (\ref{eq:PZN-1}){]} have five and
seven parameters, respectively. Ideally, one hopes to be able to reduce
this large number and use as few parameters as possible. However,
it shall be noted that, among these parameters, many are only material
dependent (i.e., they do not depend on frequency or temperature).
For instance, for lead-free relaxor, $E_{b}$ is a constant; for lead-based
relaxor, $E_{b}$, $\chi_{1}$, $\chi_{2}$, $\theta$ are close to
constants (see Tab. \ref{tab:Fitting-parameters-of-PZN}). For a given
sample, these parameters may only need to be calibrated once. In this
way, the number of parameters will be significantly reduced. 

Second, in this work we have focused on the temperature dependence
of susceptibility. The dependence on frequency needs further investigation.
For instance, $T_{C}\left(\nu\right)$ and $\gamma\left(\nu\right)$
in Eq. (\ref{eq:PZN-1}) need to be specified explicitly to address
this issue.  We note that results shown in Tab. \ref{tab:Fitting-parameters-of-PZN}
will provide clues to $T_{C}$ and $\gamma$'s dependence on $\nu$,
and eventually help finding analytical expressions for $\chi\left(T,\nu\right)$.
In addition, we generally ignored the long-range correlation of dipoles,
which is another limitation to this approach. While such long-range
correlation makes relaxor physics so rich, it will bring back Bose-Einstein
statistics and make the current formulation more complicated. To what
extent the Bose-Einstein and the Maxwell-Boltzmann distribution shall
be used for ferroelectric relaxors remains an open question.

Third, at high temperature, Curie law is observed in many ferroelectrics.
For the static susceptibility of BZT {[}Fig. \ref{fig:Fitting-susceptibility-of}{]},
this law can be recovered from the proposed equation, Eq. (\ref{eq:BZT-susceptibility}).
On the other hand, for Eqs. (\ref{eq:BZT-frequency}) and (\ref{eq:PZN}),
the Curie law cannot be directly recovered. For Eq. (\ref{eq:PZN}),
we have the asymptotic relation $\text{\ensuremath{\chi}}\sim A/\left(T-T_{C}\right)^{\gamma}+B/T^{3/2}$
at very large $T$. It is unclear how good this relation can fit the
Curie law. Therefore, in fitting experimental data at high temperatures,
one needs to bear in mind that Eqs. (\ref{eq:BZT-frequency}) and
(\ref{eq:PZN}) should be used with care. 

Finally, with Eqs. (\ref{eq:BZT-susceptibility}) and (\ref{eq:PZN}),
in principle we can obtain the relation between $T_{m}$ and $\nu$,
which can then be compared to the well-known Vogel-Fulcher law \cite{Wang2016}.
However, we have failed to obtain analytical expressions for $T_{m}\left(\nu\right)$
and believe numerical calculation seems to be the only feasible way
to establish the relation between $T_{m}$ and $\nu$.

\section{Conclusion\label{sec:Conclusion}}

Instead of working on the atomic level, the present work employs a
macroscopic statistical approach to help understanding dielectric
properties of relaxors. The effects of disorder, PNRs, and random
electric fields are considered statistically by introducing the average
potential well, which can trap dipoles of low kinetic energy. An external
electric field will mostly increase the magnitude\textbf{ }of trapped
dipoles, but rotate to align dipoles free from such trapping, giving
rise to two different types of dielectric responses as shown in Eqs.
(\ref{eq:BZT-susceptibility}), \ref{eq:BZT-frequency}, and (\ref{eq:PZN}).
This approach, by proposing analytical equations, provides insights
to experimental and numerical results of relaxors.\textbf{ }Among
other things, it shows that the characteristic temperature, $T_{m}$,
is determined by the Maxwell-Boltzmann distribution of dipoles' kinetic
energy, as well as their ability to respond to the applied electric
field. We can also conclude that lead-free relaxors (e.g., BZT) are
different from lead-based relaxors (e.g., PZN-0.13PT) in that (i)
The mechanisms determining $T_{m}$ are different. For lead-based
relaxors, it appears $T_{O}$ alone in able to determine $T_{m}$,
while for BZT, both $w_{1}\left(T\right)$ and $P\left(E_{b},T\right)$
are important; and (ii) For BZT, $\chi_{1}$ and $\chi_{2}$ are on
the same order, in contrast to the fact while $\chi_{1}\gg\chi_{2}$
for the Pb-based relaxors, indicating that dipoles outside the average
potential well dominate dielectric response of Pb-based relaxors.
With these results, we hope this statistical approach can help better
understanding important relaxor systems and the proposed equations
be adopted in fitting experimental data.

\begin{acknowledgments}
We thank Drs C.-L. Wang, A. A. Bokov and L. Bellaiche for fruitful
discussion. This work is financially supported by the National Natural
Science Foundation of China (NSFC), Grant No. 51390472, 11574246,
U1537210, and National Basic Research Program of China, Grant No.
2015CB654903. F.L. acknowledges NSFC Grant No. 51572214. Z.J. acknowledges
the support from China Scholarship Council (CSC No. 201506280055).
We also acknowledge the ``111 Project\textquotedblright{} of China
(Grant No. B14040), the Natural Science and Engineering Council of
Canada (NSERC) and the United States Office of Naval Research (ONR
Grants No. N00014-12-1-1045 and N00014-16-1-3106).
\end{acknowledgments}

\providecommand{\natexlab}[1]{#1} \providecommand{\url}[1]{\texttt{#1}}
\expandafter\ifx\csname urlstyle\endcsname\relax \providecommand{\doi}[1]{doi: #1}\else
\providecommand{\doi}{doi: \begingroup \urlstyle{rm}\Url}\fi

\renewcommand{\doi}[1]{}

\end{document}